\documentclass[12pt]{article}
\usepackage{graphicx,color,subfigure,cite}
\usepackage{hyperref}
\def\bea{\begin{eqnarray}}
\def\eea{\end{eqnarray}}
\def\ltap{\raisebox{-.4ex}{\rlap{$\sim$}} \raisebox{.4ex}{$<$}}
\def\gtap{\raisebox{-.4ex}{\rlap{$\sim$}} \raisebox{.4ex}{$>$}}
\def\beq{\begin{equation}}
\def\eeq{\end{equation}}
\def\barr{\begin{array}}
\def\earr{\end{array}}
\def\dis{\displaystyle}
\def\gev{\, {\rm GeV}}
\def\tev{\, {\rm TeV}}

\begin{document}
%\begin{flushleft}
%\today
%\end{flushleft}
\begin{center}
{\large \bf A fourth generation, anomalous like-sign dimuon charge
asymmetry and the LHC\\}
\vglue 0.5cm
Debajyoti Choudhury $^{a}$ and Dilip Kumar Ghosh$^{b}$ \\[2ex]
{$^a$ \em Department of Physics \& Astrophysics,\\
University of Delhi, Delhi - 110 007, India }\\[1ex]
{$^b$ \em Department of Theoretical Physics,\\
Indian Association for the Cultivation of Science,\\  
2A \& B Raja S.C. Mullick Road, Kolkata 700 032, India}\\
\end{center}

%\pacs{}
%\preprint{}
%%%%%%%%%%%%%%%%%%%%%%%%%%%%%%%%%%%%%%

\begin{abstract}
A fourth chiral generation, with $m_{t^\prime}$ in the
range $\sim (300 - 500)$ GeV and a moderate value of the 
CP-violating
phase can explain the anomalous like-sign dimuon charge
asymmetry observed recently by the D0 collaboration. The 
required parameters are found to be consistent with constraints 
from other $B$ and $K$ decays. The presence of such quarks,  
apart from being detectable in the early stages of the LHC, would
also have important consequences in the electroweak 
symmetry breaking sector.
 \end{abstract}
%\input{intro.sect}
%%%%%%%%%%%%%%%%%%%%%%%%%%%%%%%%%%%%%%
\section{Introduction}
Charge-parity (CP) violation is one of the key ingredients for the
dominance of matter over anti-matter in the present day universe. 
The presence of a complex phase in the Cabbibo-Kobayashi-Maskwaw 
(CKM) matrix \cite{cabibbo,koba_maska} leads to CP violation in the 
Standard Model(SM). Since the first observation of the CP violation
in the Kaon system \cite{cpv_kaon}, several
measurements have been performed to observe the same in $B$ and $D$
mesons and the experimental findings have, so far, been largely 
consistent with the
prediction of the SM. However, the SM neither admits 
sufficient CP violation nor is the phase transition 
strong enough to explain the observed magnitude of the 
baryon asymmetry of the universe. 
This has led to a sustained effort over decades to look for an 
evidence of CP-violation going beyond what the SM predicts. Indeed, 
during the past few years some experimental data collected 
at Tevatron and B-factories in the heavy flavour-sector 
indicates mild conflict with the CKM picture of CP violation within the
SM \cite{lunghi-soni123,lenz,utfit}. Several new scenarios have been proposed
to explain these anomalies\cite{agashe,blanke1,blanke2,neubert,barger,buras}.
Much interest has been garnered by the recent
observation\cite{D0_dimuon}, by the D0 Collaboration, of an anomalous
like-sign dimuon charge asymmetry. They find an excess of
negatively-charged dimuon pair over positively charged ones, namely
$ N^{--} \equiv N(\mu^- \mu^-) > N^{++}$. 
Reconstructing the events, and subtracting the backgrounds, they
conclude that the only possible explanation is offered by the
conjecture that a $b + \bar b$, created in a hard process,
hadronize into a pair of neutral $B$-mesons
(say, $B_s + \bar B_s$)
each of which suffers a semileptonic [$B({\overline B}) \to
\mu^{-}(\mu^{+}) + X $] decay. The oscillation of one into
the other allows for a `wrong-sign' decay leading to like-sign
dimuons. The asymmetry, then, would be a consequence of unequal 
probabilities of $B_s \to \overline B_s$ and $\overline B_s \to B_s$
oscillations, a manifestation of CP-noninvariance. 
Thus, the D0 measurement\cite{D0_dimuon},  using $6.1 {\rm fb}^{-1}$ of data, 
of 
\begin{eqnarray}
A^b_{sl} \equiv \frac{N^{++}_b - N^{--}_b}{N^{++}_b + N^{--}_b} = 
-(9.57 \pm 2.51 \pm 1.46)\times 10^{-3} \, 
\label{D0_Absl}
\end{eqnarray}
would amount to a $3.2\sigma $ deviation 
away from the SM prediction of $-0.2\times 10^{-3}$. On the other hand,
the CDF Collaboration \cite{cdf_dimuon}, using $1.6~{\rm fb}^{-1}$ of data
has also measured $A^b_{sl}$ with a positive central value, namely
$A^b_{sl} = (8.0 \pm 9.0 \pm 6.8)\times 10^{-3}$. Because of the large
errors in the CDF measurements, it is still compatible with the
D0 one within $1.5\sigma $ level. Combining the two, one obtains
\begin{eqnarray}
A^b_{sl}\approx -(8.5 \pm 2.8) \times 10^{-3} \, ,
\label{Absl_comb}
\end{eqnarray}
which still is $3\sigma $ away from the SM value.

It should be noted at this stage that the 
like-sign charge asymmetry $A^b_{sl}$ measured by D0 Collaboration
\cite{D0_dimuon} is related to the semileptonic decay asymmetries 
$a^d_{sl}$ and $a^s_{sl}$ in the $B_d$ and $B_s$ mesons, respectively,
through 
\begin{eqnarray}
A^b_{sl} = (0.506 \pm 0.043) \, a^d_{sl} + (0.494 \pm 0.043) \,
a^s_{sl}  \, ,
\label{Absl_Bd_Bs}
\end{eqnarray}
where
\begin{eqnarray}
a^q_{sl} \equiv \frac{\Gamma({\bar B_q} \to \mu^+ X)-\Gamma(B_q \to
\mu^- +X)} {\Gamma({\bar B_q} \to \mu^+ X)+\Gamma(B_q \to \mu^- +X)}
\qquad (q =d,s) \, .
\label{aqsl_def}
\end{eqnarray}
The D0 Collaboration 
has also directly measured $a^s_{sl}$, 
albeit with much large errors\cite{D0_single_asym} 
\beq
a^s_{sl} = -(1.7 \pm 9.1^{+1.4}_{-1.5})\times 10^{-3} \ ,
\label{assl_D0}
\eeq
whereas the current direct bound on $a^d_{sl}$ reads\cite{hfag}
\beq
a^d_{sl} = (4.7 \pm 4.6)\times 10^{-3} \ .
\label{adsl_hfag}
\eeq
At this stage, it should be realized that, within the SM, 
$a^d_{sl}$ would contribute negligibly to $A^b_{sl}$. And since the 
new physics (NP) scenario that we would be considering 
entails almost a vanishingly small contribution to this quantity, 
it is meaningful to assume a SM value for the same. Thus, combining 
all the data, we have\cite{bogdan}
\beq
a^s_{sl} \approx -(12.7 \pm 5.0)\times 10^{-3}.
\label{assl_comb}
\eeq
This result is still $2.5\sigma $ away from the SM value\cite{lenz},
namely, 
\bea
a^s_{sl}\approx 0.02 \times 10^{-3} \, .
\label{absl_sm}
\eea
While a discrepancy of this magnitude cannot be considered a
definitive discovery of NP effects, it certainly adds to the existing
list of deviations from the SM expectations seen in the $b$-sector. It,
thus, becomes interesting to consider well-motivated NP scenarios that
might lead to an coherent explanation of such anomalies while leading
itself to experimental verification in other independent
channels \cite{bogdan,kundu,axigluon,gion_isidori,ligeti,babu,desh}. 
Perhaps the simplest such extension is the postulation of a
fourth family of quarks that mix with the three known families, thereby 
altering the structure of meson-mixings, CP violation, flavour-changing 
neutral currents (FCNC) etc.~\cite{hou1,hou2,hou3,soni1,hung1,arhrib-hou_bs,
alwall,Bobrowski,soni-alok2,buras_2010,hou_ma,Eberhardt,soni-alok1}. 
A striking consequence of such a fourth 
family would be the introduction of additional phases in the analogue 
of the CKM matrix, or in other words, the existence of new Jarlskog 
invariants~\cite{jarlskog}. As some of these are no longer suppressed 
by the first generation quark masses, there is an enormous enhancement in 
the effective magnitude of CP violation available to the mechanism 
of baryogenesis~\cite{hou_baryo}. 

Although the addition of vector-representations of quarks is, in 
some sense, minimal and also suffices to explain the long-standing 
anomaly in the forward-backward asymmetry in $b$-pair production 
at LEP/SLC~\cite{c_t_w}, the introduction of such representations 
typically results in tree-level FCNCs~\cite{morrissey}. We, rather, 
choose to work with the more canonical scenario, namely a chiral 
fourth generation $(t', b')$ quark model (SM4).

The CDF collaboration has looked for the existence of such quarks and
quote bounds of $m_{t'} > 335 \gev$~\cite{cdf_tpr} and $m_{b'} > 338
\gev$~\cite{cdf_bpr} respectively. However, as both these analyses
assume a 100\% decay branching fraction into particular modes, the
model-independent bounds would be relaxed somewhat and heavy quarks of
$m_{t', b'}~\gtap~300 \gev$ are still quite consistent. On the other
hand, the introduction of chiral fermions can cause deviations in
electroweak precision test variables (in particular, the custodial
symmetry parameter $\rho$) from their SM values, and this severely
constrains the allowed mass splitting between the quarks to 
$| m_{t'} - m_{b'} |~\ltap~70 $ GeV
~\cite{polonsky,langacker,novikov1,novikov2,Kribs_EWPT,chanowitz,hashimoto}. 

The rest of this article is organized as follows. In Section 2, we discuss
the like-sign dimuon charge asymmetry from $B_s - {\bar B_s}$ mixing in 
the SM4.  We also show the constraints on the parameters space of SM4 
obtained from $b \to s \gamma $ process. In Section 3, we discuss the
possible mechanisms for a direct search of heavy $t'/b'$ quarks at the LHC. 
Finally, in Section 4, we summarize our findings.
%%%%%%%%%%%%%%%%%%%%%%%%%%%%%%%%%%%%%%
%\input{bphys.sect}
%%%%%%%%%%%%%%%%%%%%%%%%%%%%%%%%%%%%%%
\section{$ B_s - {\bar B_s}$ mixing}
The lighter $(L)$ and the heavier $(H)$ mass eigenstates of the 
$B_s$ system are split with sizeable differences for both mass 
($\Delta M_{s} \equiv M_{sH} - M_{sL} $) and
decay widths ($\Delta \Gamma_{s} \equiv 
\Gamma_{sL} - \Gamma_{sH} $). 
Within the SM, $B_s - {\bar B_s}$ mixing
is dominated mainly by the box diagrams with top quarks and $W$-boson 
circulating in the loop. With the introduction of a fourth family, and
allowing for it to mix with the known three, the 
CKM matrix is now extended to a $4\times 4 $ unitary one. As a consequence, 
additional diagrams with a $t'$ replacing one or both of the $t$-quarks
in the loop now contribute. The expression for the mass difference
is now altered to \cite{arhrib-hou_bs}
\begin{eqnarray}
\Delta M_{s} = 2 \, | M_{12}| \ , 
\end{eqnarray}
where 
 \beq
 M_{12} = \frac{G_F^2 m_W^2}{12\pi^2} m_{B_s} B_{bs} f_{B_s}^2 
 \Big\{\eta_t \lambda^2_t S_0(x_t) + 
 \eta_{t^\prime} \lambda^2_{t^\prime} S_0(x_{t'})
 + 2 {\tilde \eta}_{t^\prime} \lambda_t \lambda_{t^\prime} 
 S_0(x_t,x_{t'}) \Big\}~,
   \label{del_m12}
 \eeq
with $x_t=m_t^2/m_W^2$, $x_{t'}=m_{t'}^2/M_W^2$ and
$\lambda_{t} = V_{ts}^{*} V_{tb} $,  
$\lambda_{t^\prime } = V_{t^\prime s}^{*} V_{t^\prime b}$.  The
loop integrals $S_0(x), S_0(x,y)$ can be found in Ref.\cite{arhrib-hou_bs}. 
The quantities $\eta_t, \eta_{t'} $ and $\tilde \eta_{t'}$ 
represent the QCD corrections accrued from running the 
effective operator obtained by integrating out the heavy fields
down to the $B$-meson scale. For example, the SM 
QCD factor $ \eta_t = 0.5765\pm 0.0065 $ \cite{buras_jamin}, while 
\bea
\eta_{t^\prime} = \left [ \alpha_s(m_t) \right]^{6/23}
                   \left[\frac{\alpha_s(m_b^\prime)}
{\alpha_s(m_t)}\right ]^{6/21} \left[\frac{\alpha_s(m_t^\prime)}
{\alpha_s(m_b^\prime)}\right ]^{6/19} 
\eea
with analogous expressions for $\tilde \eta_{t^\prime}$. Owing to the 
relatively small running of $\alpha_s$ between these scales, 
numerically, $\eta_{t^\prime}, \tilde \eta_{t^\prime} \approx
\eta_t$. 

The unitarity of the CKM-4
matrix implies $\lambda_u + \lambda_c + \lambda_t + \lambda_{t^\prime} = 0$.
Neglecting $\lambda_u$ in comparison with the others (an excellent 
approximation), we may write 
\bea
\lambda_t \cong -\lambda_c -\lambda_{t^\prime} 
\eea
with $\lambda_c = V^{*}_{cs}V_{cb}$ being real by convention. For 
$V_{cb}$, we use the value quoted in Table \ref{tab:input_data}.
Parametrizing 
\bea
\lambda_{t^\prime} = r_{t'} \exp(i\phi_{t'})
\label{rsb_phisb}
\eea
where $\phi_{t'}$ is the new CP violating phase, 
we can now express all the NP effects essentially 
in terms of $m_{t'}, r_{t'}$ and $\phi_{t'}$.

\begin{table}[!ht]
\begin{center}
\begin{tabular}[ht]{lc}
\hline
parameter  & value \\
\hline
$m_t(m_t)$   & $ (163.5 \pm 1.7) $ (GeV) \\
$ \alpha_s(M_Z)$ & 0.118 \\
$V_{cb} $ & $(40.8 \pm 0.6)\times 10^{-3} $ \\
${\rm BR}(B\to X_s\gamma) $ &  $ (3.55 \pm 0.25)\times 10^{-4}$ 
\cite{hfag}\\
${\rm BR}(B \to X_c\ell \nu) $ & $(10.61\pm 0.17)\times 10^{-2}$
\cite{soni-alok2}\\
$ f_{bs}\sqrt{B_{bs}} $ & $ (0.275 \pm 0.013) $ GeV \cite{buras_2010}\\
\hline
\end{tabular}
\caption{Values of different input parameters used in this analysis.}
\label{tab:input_data}
\end{center}
\end{table} 

%Within the SM, the effective CP violating phase $\Phi_s $ associated with 
%$B_s$--${\bar B_s}$ mixing is doubly 
%Cabibbo suppressed and  can be expressed as \cite{lenz,p_ball}
%\beq
%\barr{rclcl}
%\Phi^{\rm SM}_{s} &= & \dis 
%-2\beta^{\rm SM}_{s} & = & \dis - 2\lambda^2 \eta \, , 
%\\[2ex] 
%\beta^{SM}_{s} &\equiv & \dis \beta^{J/\psi\phi(\rm SM)}_{s} & = & \dis
%Arg\left[-\left( \frac{V^*_{tb}V_{ts}} {V^*_{cb}V_{cs}}\right )
%\right ]\approx 0.019 \pm 0.001 \, ,
%\earr
%\label{sm_phase}
%\eeq
%where $\lambda $ and $\eta $ are the usual Wolfenstein parameters of the 
%CKM-3 matrix, and $2\beta^{J/\psi\phi(\rm SM)}_{s}$ is the angle which controls
%the mixing  induced CP asymmetry in $B_s \to J/\psi \phi $ decay. 
%Furthermore, 
Within the SM, 
\bea
\Delta M_{s}^{\rm SM}  & = & (19.82 \pm 1.87)~{\rm ps}^{-1}\\
\Delta \Gamma_{s}^{\rm SM}  & = & (0.096 \pm 0.039)~{\rm ps}^{-1}
\label{Bs_theo}
\eea
where the theoretical uncertainty in $\Delta M_{s}^{\rm SM}$ 
arises mainly from the 
uncertainty in the combination 
$f_{bs}\sqrt{B_{bs}}$ (see Table \ref{tab:input_data}) of the
$B_s$ decay constant and the bag parameter. Consequent to the 
remarkable sensitivity of  both the D0 and CDF experiments,
the average decay width $\Gamma_{s}\equiv 
(\Gamma_{sL} + \Gamma_{sH})/2$ and the mass difference $\Delta M_s $
have been measured with an accuracy better than $2\%$ 
\cite{Amsler,D01,CDF1,utfit,kundu} and we now have
\beq
\barr{rcl}
\Gamma_{s} & = & \dis 1.472^{+0.024}_{-0.026}~~~~{\rm ps}^{-1}\\[1.5ex]
\Delta M_{s} & = & 17.77 \pm 0.10 \pm 0.07 ~~~~{\rm ps}^{-1} \, .
\earr
\label{Bs_data}
\eeq
On the other hand, measurements done at CDF and D0 experiments 
\cite{Aaltonen1,Aaltonen2} 
using the technique of angular analysis in 
$B_s \to J/\psi + \phi \; (b \to s c {\bar c} ) $ 
decay \cite{dighe1,dighe2}, provide \cite{hfag}
\beq
\barr{rcl}
\Delta \Gamma_{s}  & = &  \dis
\pm ( 0.154^{+0.054}_{-0.070})~{\rm ps}^{-1}
\\[1.5ex]
\beta^{J/\psi \phi}_s  & \in &  (0.39^{+0.18}_{-0.14})\cup 
(1.18^{+0.14}_{-0.18}) \, ,
\earr
\label{bs_mixing}
\eeq
where the last line reflects the two-fold ambiguity 
 ($\beta_s^{J/\psi \phi} \leftrightarrow \pi - \beta_s^{J/\psi \phi}$)
in the experimental determination of 
$\beta_s^{J/\psi \phi}$. Note also that only the
magnitude of $\Delta \Gamma_{s}$ is determined, and not the sign.

Finally, the semileptonic decay asymmetry $a^s_{sl}$ as defined in 
eq. (\ref{aqsl_def}) can be related to the $B_s$--${\bar B_s}$ mixing 
parameters, {\em viz.},
\bea
\tan\phi_{s} = \chi_s  \equiv a^s_{sl} \; 
	     \frac{\Delta M_{s}} {\Delta \Gamma_{s}} 
\label{asl_rsb_phi_sb}
\eea
where $\phi_{s} $ contains both the SM phase $(\sim 0)$ 
and the NP phase $\phi_{t'}$. 

The above equation is, in some sense, the master 
formula for our analysis. Of the 
three quantities defining $\chi_s$,
the mass difference $\Delta M_s$ has been measured very accurately 
(\ref{Bs_data}). On the other hand, the large error bars in 
$\Delta \Gamma_s$, combined with the ambiguity in its sign, allows
for large (and disjoint) swathes of $\chi_s$.

Note that there exists some tension between 
$\Delta \Gamma_{s}^{\rm SM}$ and $\Delta \Gamma_{s}^{\rm exp}$, 
although they cannot yet be deemed to be inconsistent with each other. 
The difference could be attributable to theoretical uncertainties in the
estimation of $\Delta \Gamma_{s}^{\rm SM}$ from the higher order 
corrections as well as from the long distance effects, as also to experimental
ones. However, if the central values are given any credence, this 
apparent difference would have profound consequences. 
The width difference is given by 
\bea
\Delta \Gamma_s = 2 \left| \Gamma_{12s}\right| \cos\phi_s
\eea
where, $\phi_s \equiv Arg(-M_{12s}/\Gamma_{12s})$ and $\Gamma_{12s}$ is the
absorptive part of the $B_s - {\bar B_s}$ mixing amplitude. Within the SM 
\cite{lenz}
\bea
\phi_s = 0.0041 \pm 0.0007 
\label{sm_phase}
\eea
and, therefore, 
$ \Delta \Gamma_s^{\rm SM} \approx 2 \left| \Gamma_{12s}\right| $.
It is worth emphasizing that in SM4, there is no room for additional
contribution to $\Gamma_{12s}$
which implies that $\Delta \Gamma_s \leq \Delta \Gamma_s^{\rm SM}$.
This is true not only for SM4, but for all theories wherein no additional 
absorptive parts appear. In other words, 
any new physics scenario that seeks to enhance $a^s_{sl}$ 
without enhancing $\Gamma_{12s}$ runs the risk of rendering 
$|\sin\phi_s | \geq 1$. Thus, the only way, for all such theories, 
to explain the D0 result would be to appeal to the theoretical 
uncertainties in $\Delta \Gamma_s^{SM}$ determination and rather 
assume that the experimentally measured value reflects the truth 
accurately. This is the approach we adopt and 
with this, we may now proceed to find solutions 
in the three-dimensional parameter space ($m_{t'}, r_{t'}, \phi_{t'}$) 
that are consistent both mathematically and with the data. 

\begin{figure}[!ht] 
 \begin{center}
 \vspace*{-5ex}
 \includegraphics[width=8cm,height=9.8cm,angle=0]{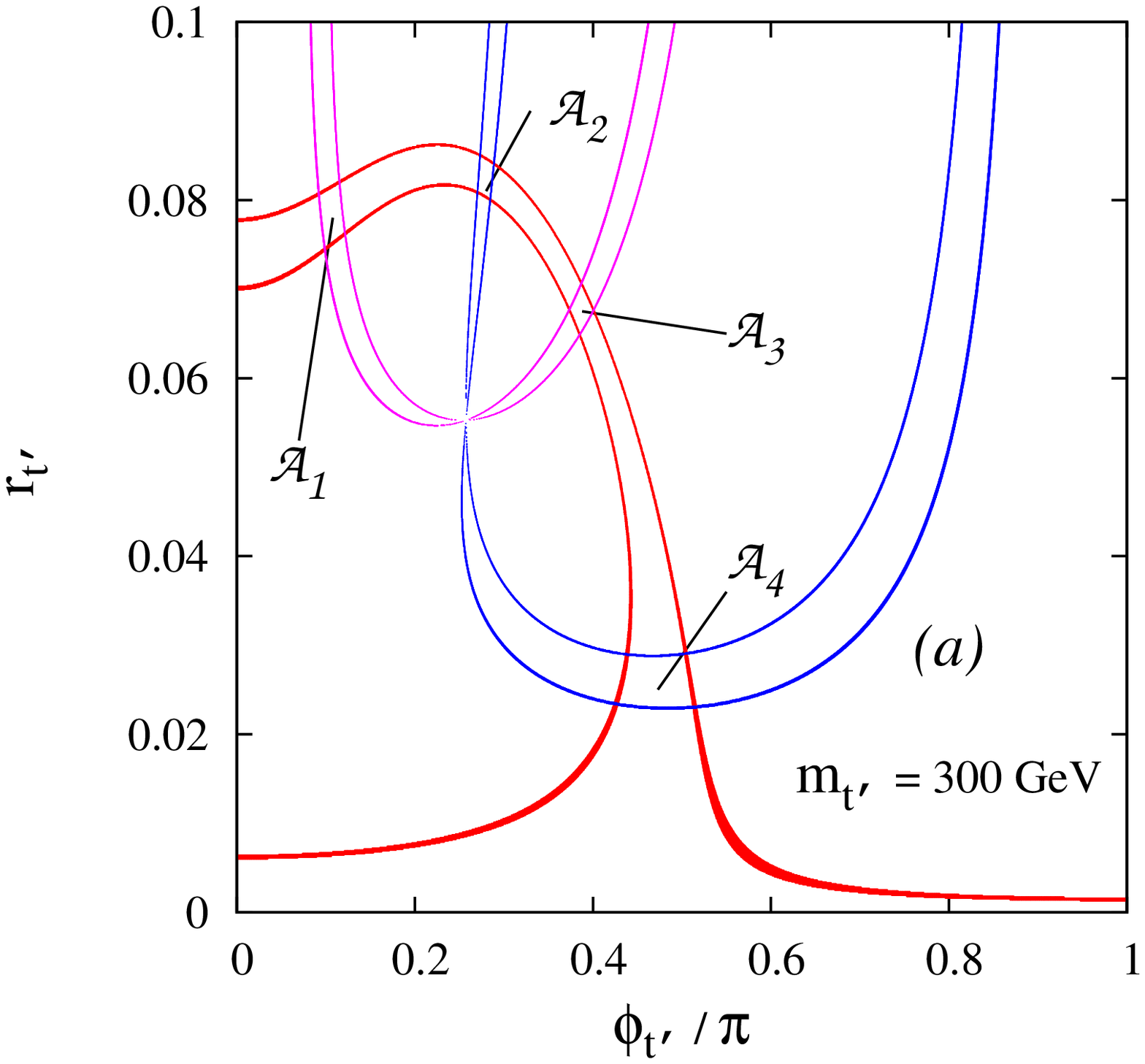}
 \hspace*{0cm}
 \includegraphics[width=8cm,height=9.8cm,angle=0]{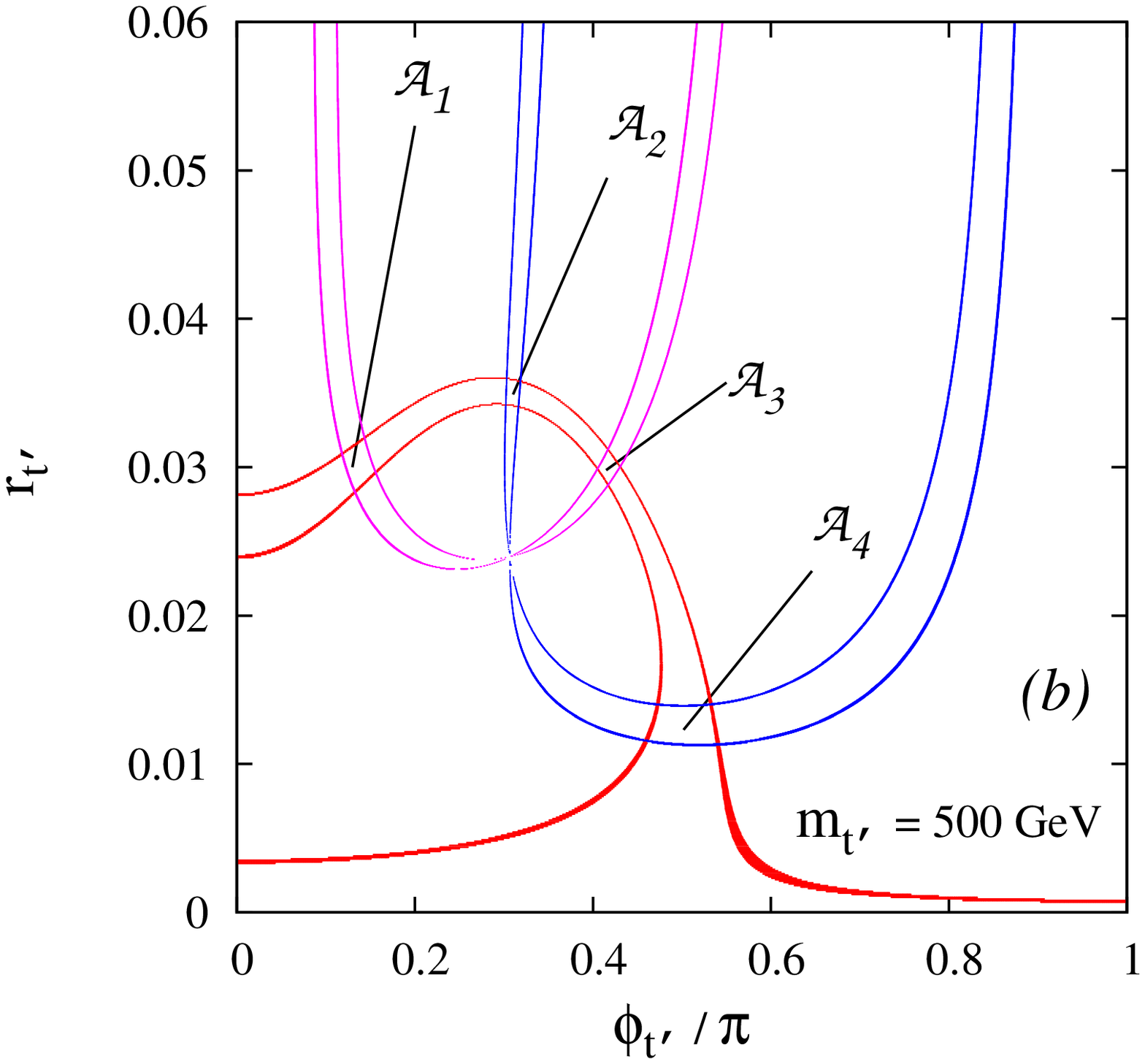}
  \end{center}
 \vspace*{-19ex}
 \caption{\em Parts of the parameter space allowed 
at $1\sigma $ by eq.\ref{asl_rsb_phi_sb} are given by the four upward 
convex strips enclosed by red or blue lines. The 
range allowed by $\Delta M_s$ is denoted by the downward 
convex strip (enclosed by green lines). The four 
areas ${\cal A}_{1,2,3,4}$ are allowed by both observables. 
The two panels correspond to different values of $m_{t'}$.}
 \label{fig:delm_asl}
 \end{figure}

Before we do so, though, it is useful
to consider an approximate symmetry in the problem. 
Given the fact
that $\phi_s^{SM}$ is a small quantity (see eq.\ref{sm_phase}), 
in the limit of neglecting it, the allowed parameter space would have
a $\phi_{t'} \leftrightarrow 2 \, \pi - \phi_{t'}$ symmetry. We may,
thus, restrict ourselves essentially to $\phi_{t'} \in [0, \pi]$. 
Having done this, for a given combination of ($\chi_s, m_{t'},
r_{t'}$), eq.\ref{asl_rsb_phi_sb} turns out to be quartic one in 
$\cos\phi_{t'}$ leading to four solution families. The inclusion of 
experimental errors would expand each curve into a band, and in 
Fig.\ref{fig:delm_asl} we display the $1\sigma$ allowed regions in 
the $\phi_{t'}$--$r_{t'}$ plane for two representative values of 
$m_{t'}$. Only the (four) strips, each enclosed between a pair 
of  similar curves, are allowed. As expected, the origin
(corresponding essentially to a fourth generation decoupled from 
the known three) lies clearly
outside the allowed region. 

This parameter space is further constrained by the precise measurement 
of $\Delta M_s$. Solving eq.\ref{del_m12} for a given $m_{t'}$, we 
once again get a $1\sigma$ allowed band in the $(\phi_{t'},r_{t'})$ 
plane (see Fig.\ref{fig:delm_asl}). Now, of course, the origin is 
included in the allowed set. The intersection of the two sets, then, 
gives us the allowed region consistent (within $1\sigma $) 
with the experimental measurements
of both the single charge asymmetry parameter $a^s_{sl}$ as well as 
$\Delta M_s$. Note that the said intersection is a
union of four disjoint areas of the parameter space, denoted in 
Fig.\ref{fig:delm_asl} by ${\cal A}_{1,2,3,4}$. 
It has been argued\cite{soni-alok1, chanowitz} that the measurement of
$Z \to b \bar b$ at LEP and SLC puts very strong constraints
on $r_{t'}$ \footnote{The values of $m_{t'}, r_{t'}$ and $\phi_{t'}$ considered
in \cite{soni-alok1} are consistent with constraint from $K^+ \to \pi^+ \nu \nu$
decay.}. If we take this at face value, then, of the four 
disjoint areas, only part of ${\cal A}_4$ remains. 
However, apart from the fact that the 
constraints of Ref.\cite{soni-alok2} have been obtained with a 
slightly differing set of inputs and cannot be imposed directly, 
note that that both our allowed ranges as well as 
the bounds of Ref.\cite{soni-alok2} are at $1\sigma$ and, even at  
only $1.5 \sigma$, the overlap is quite extensive. Furthermore, the
said strong bounds do hinge upon the assumption of no other new
physics being of relevance. In view of this, we would advocate that 
all the four regions be considered seriously and the final outcome 
be decided by further analyses, both experimental and theoretical.

We now consider possible constraints on the
NP parameter space from the measurement of the 
inclusive $b \to s \gamma$ transition, namely 
${\rm B} (B \to X_s \gamma )$. 
This one-loop process, 
in the SM, is dominated by the diagram involving a 
virtual top quark and $W$ boson. The not-so-inconsiderable 
branching ratio, combined with the theoretical cleanliness 
make this an ideal theater for testing any new theory bearing on flavour. 
In the presence of the fourth generation, additional 
contribution to the 
$b \to s\gamma $ amplitude would accrue from the 
$t^\prime $--loop. 

The effective Hamiltonian for the $ b \to s \gamma $ process can be
written as
\beq
{\cal H}_{eff} = \frac{4 G_F}{\sqrt{2}} \; V^{*}_{ts}V_{tb} \; \sum_{i=1}^{8}
C_i(\mu) \, Q_i(\mu).
\eeq
where the operators $O_i(\mu)$ and the Wilson coefficients 
$C_i(\mu)$ may be found in Ref.\cite{buras_munz}. 
The fourth general manifests itself essentially in the modification 
of the two Wilson coefficients $C_7$ and $C_8$~\cite{soni-alok2}, namely
\beq
C^{\rm tot}_{7,8}(\mu) = C_{7,8}^{SM}(\mu) + \frac{V_{t's}^{*} V_{t'b}} 
{V_{t s}^{*} V_{tb}} \; C^{t'}_{7,8}(\mu)
\eeq
where the new contributions 
$C^{t'}_{7,8}$ can be obtained from $C^{SM}_{7,8}$ simply by 
replacing the top quark mass $m_t$ in the latter by $m_{t'}$. 
There exists 
a large uncertainty 
in the estimation of the different Wilson coefficients 
due to the definition of the bottom quark mass.  This uncertainty
can be reduced by considering, instead, the ratio \cite{soni-alok2}
\beq
R = \frac{{\rm BR}(B \to X_s \gamma)}{{\rm BR}(B \to X_c e {\bar \nu_e})}
\eeq
where $ {\rm BR}(B \to X_c e {\bar \nu_e}) $ represents the semi-leptonic
branching ratio of the $B$ meson into charmed states.
In the leading logarithmic approximation, the ratio $R$ can be
conveniently expressed as 
\beq
R = \frac{\mid V^{*}_{ts}V_{tb}\mid^2}{\mid V_{cb}\mid^2 } 
\frac{6\alpha \mid C^{\rm tot}_{7}(\mu)\mid^2}{\pi f(x_c) \kappa(x_c)}
\eeq
where, $x_c \equiv m_c/m_b$ and the phase space factor $f(x_c)$ and
the QCD correction factor $\kappa(x_c)$ 
can be found in Refs.\cite{soni-alok2, nir_1989}.

\begin{figure}[htb] 
\begin{center}
\vspace*{5ex}
\includegraphics[width=8cm,height=9.8cm,angle=0]{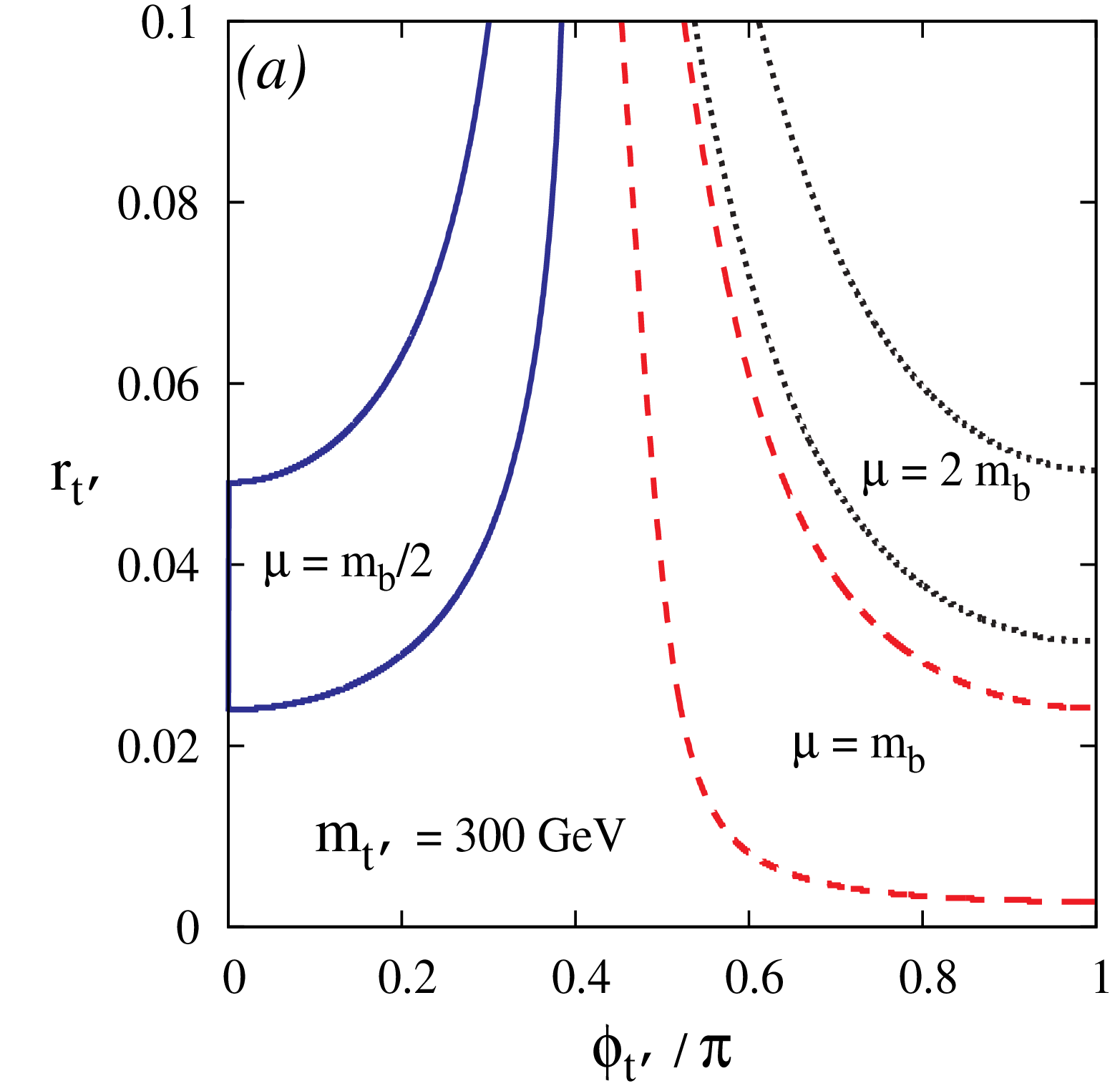}
\hspace*{0cm}
\includegraphics[width=8cm,height=9.8cm,angle=0]{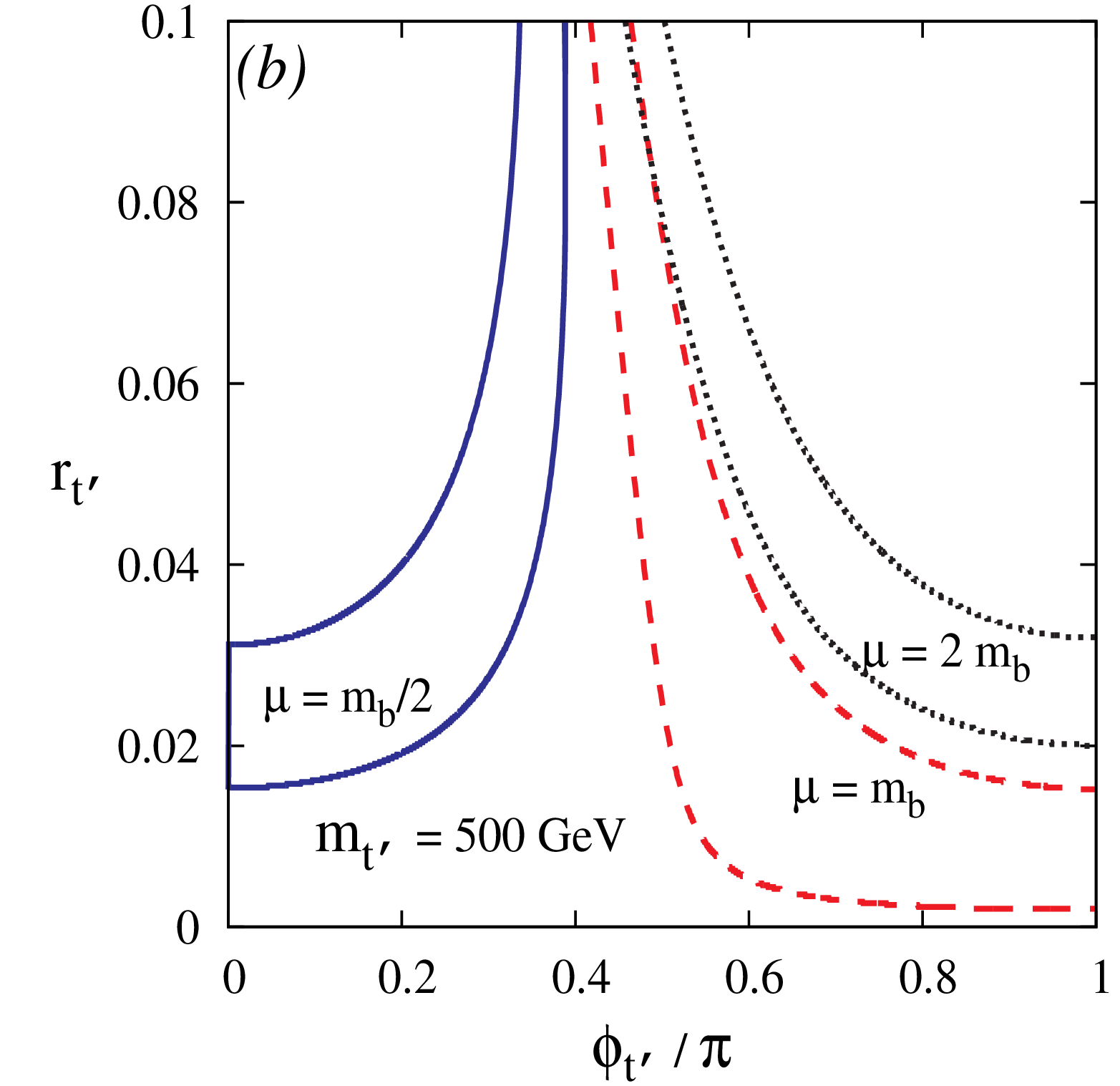}
 \end{center}
\vspace*{-22ex}
\caption{\em The $1\sigma $ allowed bands for the ratio
${\rm B}(B \to X_s\gamma)/ B(B \to X_c e \bar \nu_e)$ 
in the $(\phi_{t'}, r_{t'})$ plane for two sample choices of
$m_{t'}$. 
The dependence on the choice of the scale $\mu$ is also displayed.}
\label{fig:bsg_cont}
\end{figure}

In Fig. \ref{fig:bsg_cont}, we display the $1\sigma$ allowed 
bands for the ratio $R$ in the $r_{t'}$--$\phi_{t'}$ plane 
for two representative choices of $m_{t'}$. 
Note that there is a very strong dependence on the 
choice of the scale $\mu$. 
Given the uncertainty in this 
choice, it is thus impossible to further constrain the allowed 
parameter space using this data (as can be 
easily ascertained by a comparison of 
Figs. \ref{fig:delm_asl} and \ref{fig:bsg_cont}). 
For a give value of the scale $\mu$, it is obvious that 
a heavier $t^\prime$ implies a smaller $r_{t'}$. 
This is quite analogous to what we also observe in
Fig.\ref{fig:delm_asl} and is easy to understand. 
Both $\Delta M_s$ and $C_7^{\rm tot}$ receive a large positive 
contribution from 
the $t'$ loop, with the contribution 
growing\footnote{The decoupling theorem does not hold here, since for large
$m_{t'}$, it couples to very strongly with longitudinal mode of the
$W$ boson.} with $m_{t'}$.
Thus, to compensate for such large enhancements, $r_{t'} $ 
must become smaller. Indeed, it is the same effect that led 
Ref.\cite{chanowitz} to conclude that electroweak precision tests
allow for a smaller $r_{t'}$ for larger $m_{t'}$. 
%%%%%%%%%%%%%%%%%%%%%%%%%%%%%%%%%%%%%%
%\input{lhc.sect}
%%%%%%%%%%%%%%%%%%%%%%%%%%%%%%%%%%%%%%
\section{LHC Signals}
It is instructive, at this stage, to consider $t' \, (b')$ signals at
the Tevatron and the LHC~\cite{HoldomYan,Bholdom,Arik,Ozcan,Saavedra,Burdman,Skiba}. Pair production of such quarks is
overwhelmingly a pure QCD process (with the $q \bar q$ initial state
dominating at the Tevatron and the $g g $ state at the LHC) and is
analogous to that of top-production, the analytic expressions for
which can be found in Ref.\cite{Combridge:1978kx}. The decay
processes, though, depend on the mass splittings and the magnitudes of
the CKM-4 matrix elements. Several channels are of interest here.  For
$m_{t'} > m_{b'}$, the $t'$ dominantly decays into $b' + f_1 + \bar
f_2$ with $f_i$ being the SM fermions that are kinematically
allowed. The only exception to this would be the case where $t'$ and
$b'$ are closely degenerate and/or the off-diagonal coupling $V_{t'
b}$ is large.  The $b'$ decay is more parameter dependent. In general,
the dominant decay mode would be $b' \to q + W$ where $q$ is the quark
with the dominant off-diagonal coupling with the $b'$ (nominally, the
top).  Again, the exception is provided by the case where $b'$ is
closely degenerate with the top-quark. In such cases, the
loop-mediated decays into $b + Z/H$ could compete with that into $u/c
+ W$~\cite{arhrib_hou_bpdk}. 
On the other hand, for $m_{t'} > m_{b'}$, the $b'$ would like to
decay into the $t'$ and two soft fermions, unless the $b'$ and $t'$
are closely degenerate, in which case the $t'$ would be replaced by
the $t$. As for the $t'$, its decay would now be almost overwhelmingly
into $b + W$ leading to top-like events\cite{Choudhury:2009kz} but 
with some differences in the  kinematical distributions. Issues 
such as this have been studied extensively in the context of the Tevatron
both experimentally\cite{cdf_tpr,cdf_bpr} and theoretically\cite{Flacco:2010rg}.

\begin{figure}[htb] \begin{center}
\vspace*{5ex}
\includegraphics[width=7cm,height=9.0cm,angle=0]{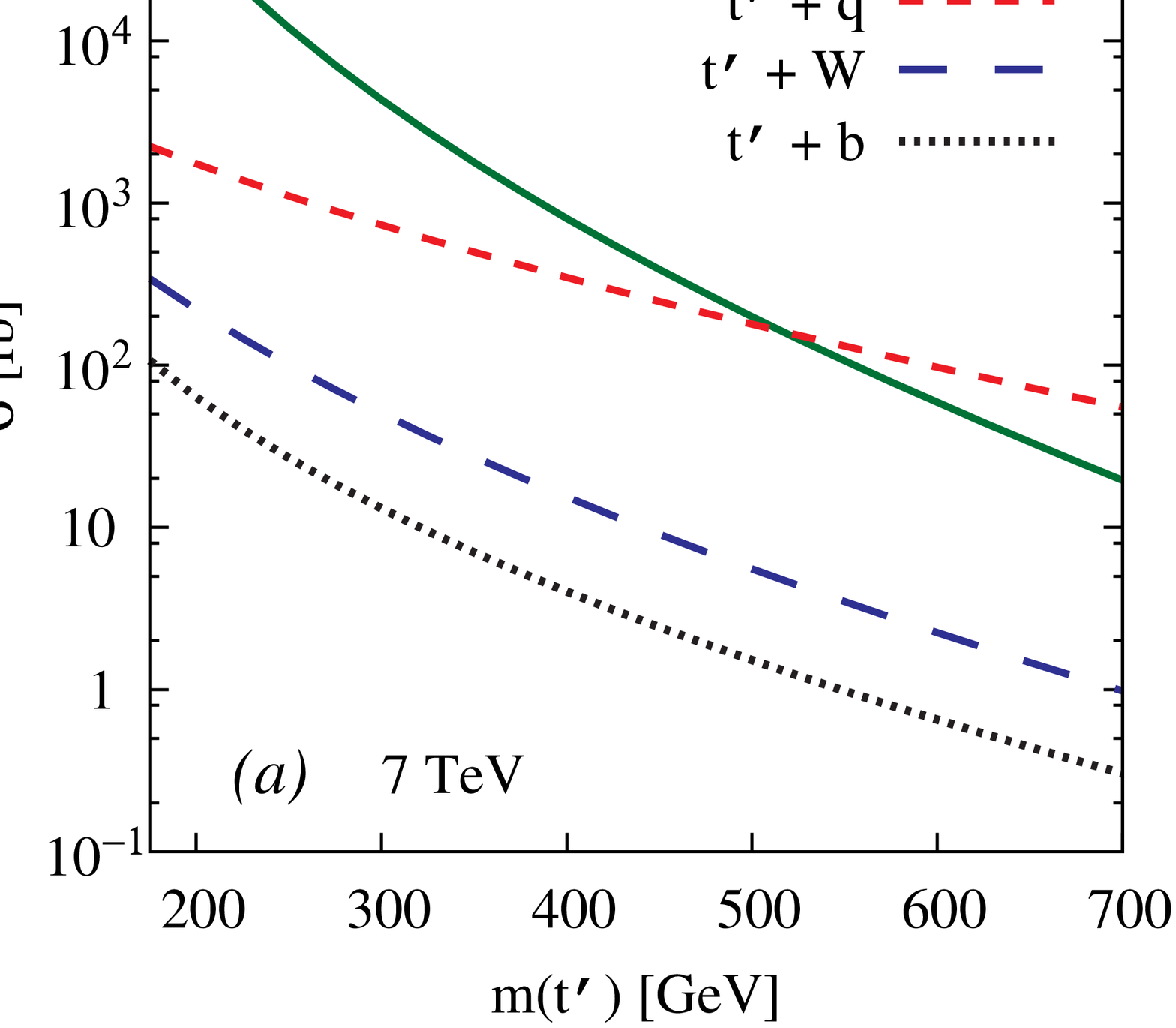}
\hspace*{0cm}
\includegraphics[width=7cm,height=9.0cm,angle=0]{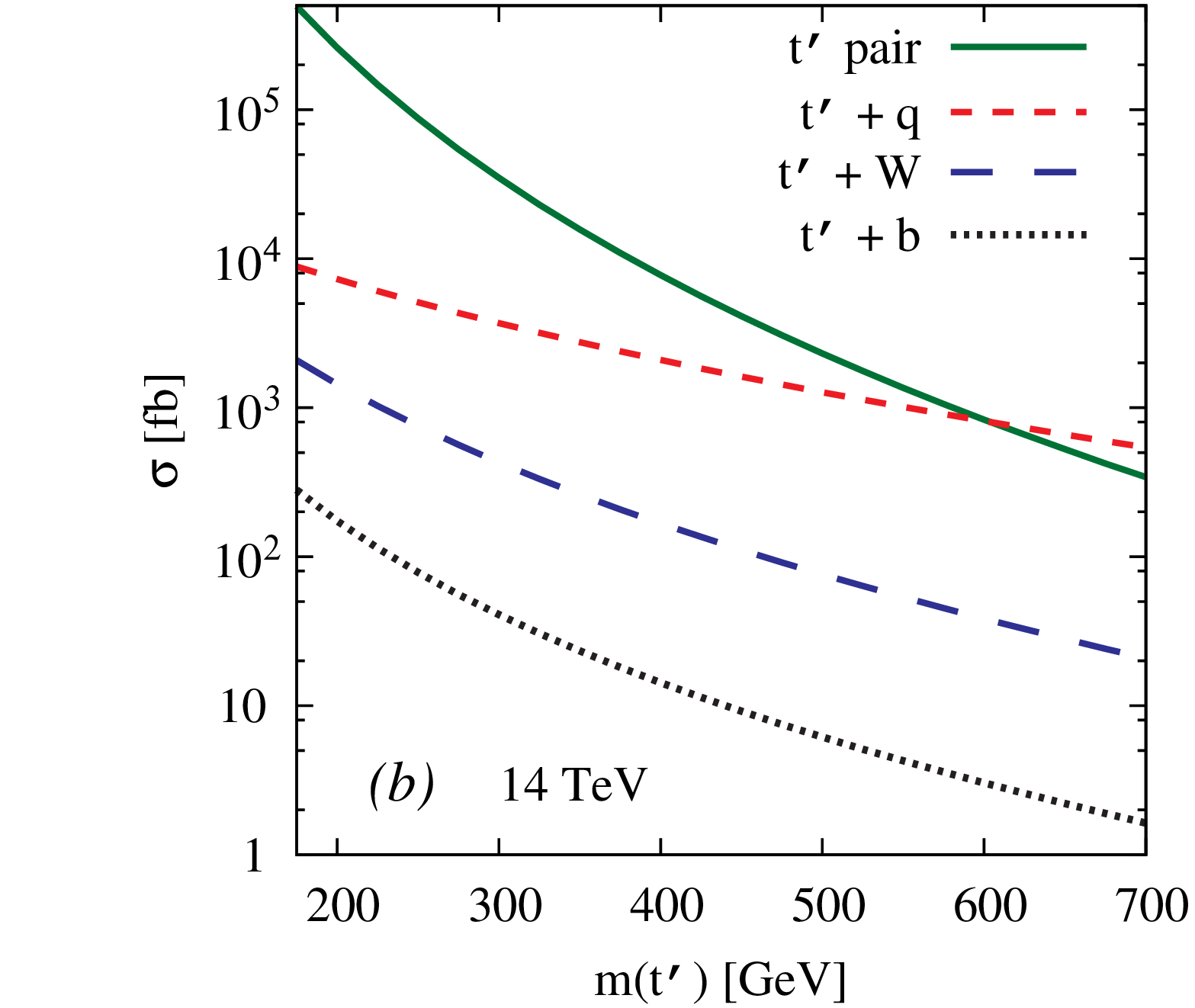}
 \end{center} 
\vspace*{-18ex}
\caption{\em Variation of $t'$ production cross sections
in different modes as a function of its mass for {\em (a)} $\sqrt{s} = 7 \tev$
and {\em (b)} $\sqrt{s} = 14 \tev$. In each the solid (green), 
short-dashed (red), long-dashed (blue) and dotted (black) curves refer 
to $t' \bar t'$, $t' + \mbox{light}-q$, $t' + W$ and $t' + b$ modes 
respectively. All cross sections are at the leading order and computed with
CTEQ6L parton densities. The weak cross sections scale with $|V_{t'b}|^2$ 
and have been computed with $|V_{t'b}|^2 = 0.04$.} 
\label{fig:cross-sect}
\end{figure}

Turning to the LHC, we display, in Fig.~\ref{fig:cross-sect}, the cross
sections for pair-production of a heavy quark $Q$ (whether $t'$ or
$b'$) as a function of its mass. While the calculation for the higher
order corrections to $t \bar t$ production\cite{ttbar} could be
adapted for this case, we desist from doing so. Note that the
pair-production could also be accompanied by one or more hard
jets. Inclusion of such processes would enhance the cross-section by
about 40--50\% almost independent of the heavy-quark
mass\cite{Choudhury:2009kz}. Since signal and the background
(typically dominated by $t \bar t$ production) are enhanced in a
similar fashion, the inclusion of such events would be expected to
increase the statistical significance. This is also helped by the fact
that the ISR for $Q \bar Q$ production would, typically, tend to be
harder than that for $t \bar t$~\cite{Plehn:2005cq, Alwall:2008qv}.

%%%%%%%%%%%%%%%%%%%%%%%%%%%%%%%%%%%%%%%%%%%%%%%%%%%%%%%%%%%%%%%%%%%

%%%%%%%%%%%%%%%%%%%%%%%%%%%%%%%%%%%%%%%%%%%%%%%%%%%%%%%%%%%%%%%%%%%

Perhaps of equal interest is the weak production of these quarks. It
is well-known that, at the LHC, single-production of the top-quark is
quite comparable to the QCD-driven pair-production, the smallness of
the weak coupling being nearly compensated for by the larger
phase-space, enhanced flux and the dynamics. Indeed, such a production
mode is of great interest both at the Tevatron and the LHC on account
of it being a direct probe of $V_{tb}$. 

A similar effect occurs here too. However rather than consider the
Cabibbo-unsuppressed process (driven by $V_{t' b'}$), we consider the
Cabibbo-suppressed processes driven by $V_{t' b}$. Note that a
hierarchy similar to that present in the CKM-3, coupled with $|V_{t'
b} \, V_{t' s}^*| \sim 0.02$ would typically mean $|V_{t' b}| \gtap
0.2$. Indeed, a large class of models~\cite{Froggatt:1978nt,
Buras:2010pi} predict that $|V_{t' b}| \sim |V_{u d}|$. In 
Fig.\ref{fig:cross-sect}, we also present the cross sections 
for three different weak sub-processes, computed with 
$|V_{t'b}|^2 = 0.04$. It should be understood that these are leading 
order results and would, in general, suffer considerable higher order 
corrections. Nonetheless, it is instructive to note that, for large 
$t'$-masses, the weak processes are comparable with, or even dominate,
the strong-production process. With the final states being quite 
different, only a detailed analysis of the corresponding background 
can tell us about the experimental viability of this mode. 

It might be argued at this stage that a very heavy fourth generation 
with considerable mixing with the third family is disfavoured 
by the electroweak precision tests~\cite{chanowitz}. This would render 
irrelevant our observation of single $t'$-production being important
for large $m_{t'}$ values. Note, however, that such observations 
are contingent upon the 4th family being the only NP source close to
the electroweak scale. If this assumption is relaxed, the constraints
do change considerably. Direct observation, or the lack of it, 
would consist the best test.It should be realized, nonetheless, that 
a very large value of the quark masses would imply a large Yukawa 
coupling, bordering on nonperturbativity. 

%%%%%%%%%%%%%%%%%%%%%%%%%%%%%%%%%%%%%%%%%%%%%%%%%%%%%%%%%%%%%%%%%%%%%
\begin{figure}[htb] 
\begin{center}
\vspace*{5ex}
\includegraphics[width=7cm,height=9.0cm,angle=0]{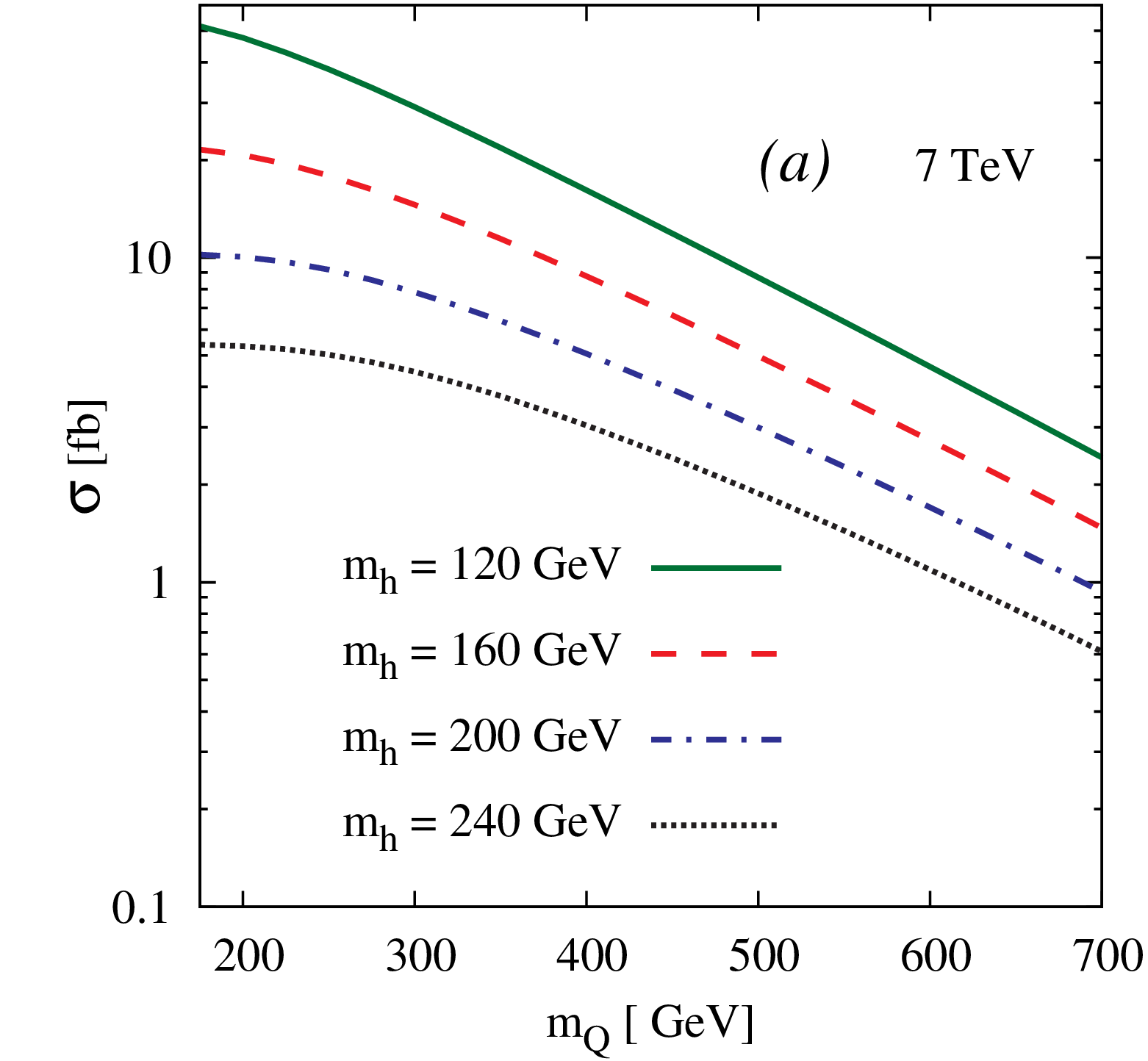}
\hspace*{0cm}
\includegraphics[width=7cm,height=9.0cm,angle=0]{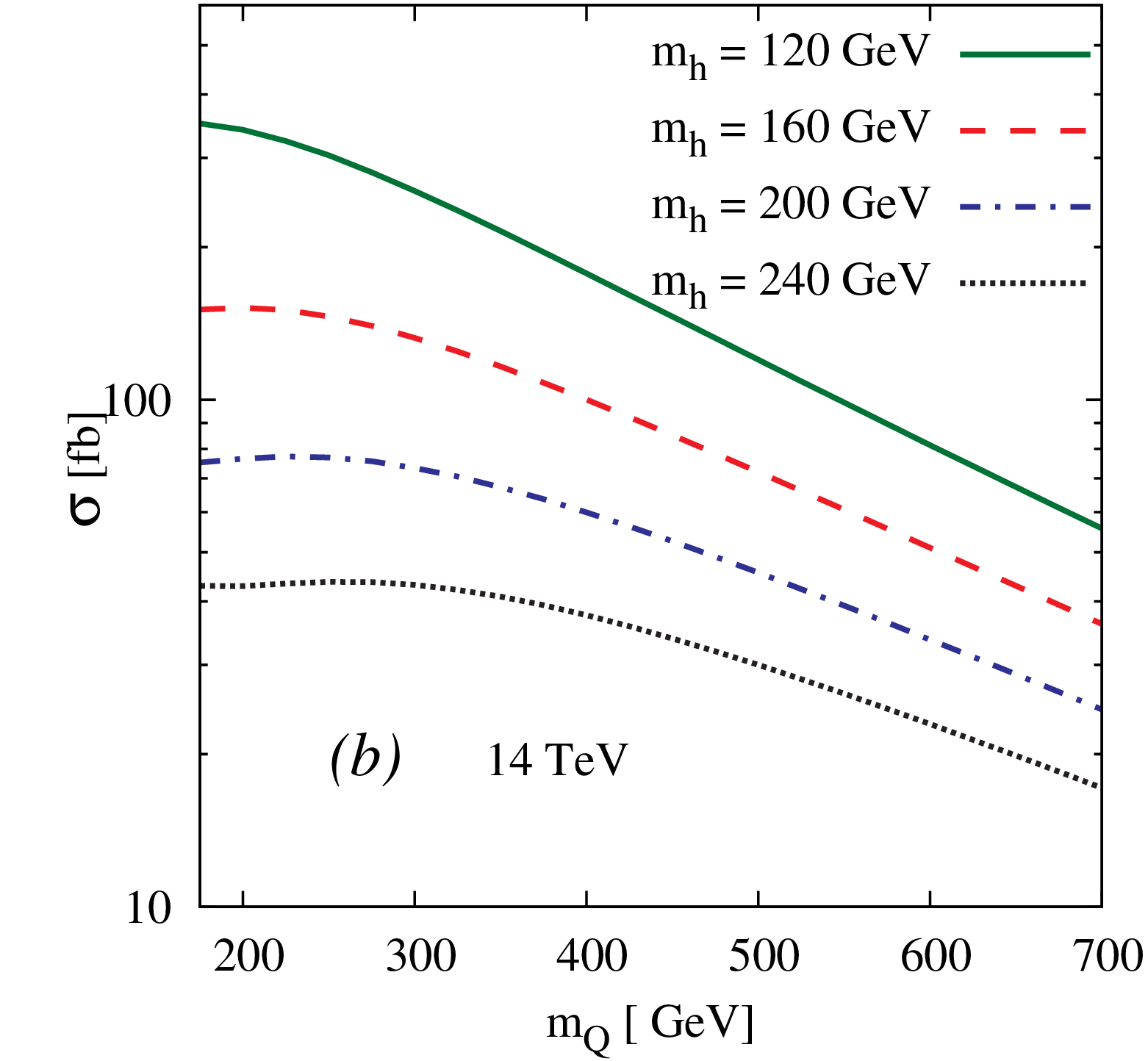}
 \end{center} 
\vspace*{-18ex}
\caption{\em The $Q \bar Q h $ production cross section at the LHC 
as a function of $m_Q$ for {\em (a)} $\sqrt{s} = 7 \tev$
and {\em (b)} $\sqrt{s} = 14 \tev$. In each the solid (green), 
short-dashed (red), long-dashed (blue) and dotted (black) curves refer 
to $m_h = 120,160,200,240 $ GeV respectively. All cross sections are at 
the leading order and computed with CTEQ6L parton densities.} 
\label{fig:tth_cross-sect}
\end{figure}
%%%%%%%%%%%%%%%%%%%%%%%%%%%%%%%%%%%%%%%%%%%%%%%%%%%%%%%%%%%%%%%%%%%

A further consequence of the existence of such heavy quarks, and
one almost independent of the magnitude of $V_{t'b}$ is the 
associated production of the Higgs boson. The process $p p \to t \bar
t h + X$ has been well-studied in the context of the LHC and both
both ATLAS\cite{ttbh_ATLAS}  and CMS collaborations\cite{ttbh_CMS}
have run extensive simulations. While initially it seemed 
that, experimentally, the channel was of marginal relevance, 
a recent reanalysis~\cite{Plehn} argues otherwise and, indeed, 
advocates its use to even measure the top Yukawa coupling. 
With the addition of the fourth
family, both $p p \to t' \bar t' h + X$ and $p p \to b' \bar b' h + X$ 
become relevant. While the higher masses of the quarks would cause 
kinematic suppression, they also imply an enhanced Yukawa coupling. 
Although the former effect does win (see Fig.\ref{fig:tth_cross-sect})
the suppression in the cross section with $m_Q$ is not very steep. 
Indeed, for $m_{t'} \ltap 500 \gev$ (the preferred range, as discussed 
above), the two modes above, together, lead to a sizable increase in
the associated Higgs production cross section, rendering it a very 
interesting mode at the LHC. It should also be realized that the
presence of such quarks would enhance the $ g g \to h$ cross section 
as well \cite{Kribs_EWPT}. 
While, for a light Higgs, the experimentally important 
two-photon decay mode also suffers a change, the last effect is not 
relevant for $m_h \gtap 160 \gev$. 
%%%%%%%%%%%%%%%%%%%%%%%%%%%%%%%%%%%%%%
%\input{concl.sect}
%%%%%%%%%%%%%%%%%%%%%%%%%%%%%%%%%%%%%%%%%%%%%%%%%%%%%%%%%%%%%%%%%%%%%%%%%%%%%
\section{Summary}
To summarise, we have sought to explain the recently claimed evidence
for an anomalous asymmetry in like-sign dimuon events by the D0
Collaboration\cite{D0_dimuon} in terms of a possible four-generation
extension of the Standard Model.  While a degenerate fourth-family is
protected from both electroweak precision tests as well as tree-level
FCNCs, the inclusion of such extra quarks, immediately leads to the
possibility of additional quark-mixing, and, hence, to additional
sources of CP-violation. Assuming, for simplicity, that the mixing of
the 4th generation with the first one is negligibly small, allows one
to significantly reduce the number of relevant new parameters,
essentially to the mass of the $t'$ and $b'$ (which need to be
relatively degenerate to protect deviations from custodial symmetry),
and the magnitude and phase of $\lambda_{t'} \equiv V_{t' b} \, V_{t'
s}$.  Confronting such a model with the experimental data, we find
that the measured value of the single-lepton decay asymmetry
$a^s_{sl}$ and the mass difference $\Delta M_s$ impose complementary
constraints on the parameter space. For a given $m_{t'}$, only four
narrow regions in the mixing parameter space are
allowed. 

If no other new physics effects are around the corner, then 
the LEP and SLC data on $Z \to b \bar b$ would disfavor two 
of the allowed regions. However, any such new physics effects and/or
a shift in the values of the hadronic parameters (such as 
$B_{b s} \, f_{bs}^2$) would change this conclusion to a significant degree.
%{\bf Interestingly, the information from $B \to K \nu \bar\nu$
%data serves to rule out two of these, leaving behind solutions 
%with a relatively small phase of $\lambda_{t'}$. Is this still true
%??}
As for $b \to s \gamma$
transitions, the solutions found herein are consistent with the data.
Given the large QCD uncertainties and the strong dependence on the scale 
at which the Wilson coefficients are calculated, the constraints
from this arena turns out to be not so crucial.

With a not too-heavy $t'$ being favoured by the data, the prospects
for detection at the LHC are very good. What is particularly
interesting is that the preferred value of $\lambda_{t'}$ indicates
that single production of $t'$-quarks may be an interesting channel to
consider.  And while such modes do win over the QCD-driven process for
very high $m_{t'}$ values, it should be noted that such large Yukawa
couplings would tend to make the theory non-perturbative relatively
early, thereby necessitating the introduction of other new physics, an
eventuality also indicated by the electroweak precision tests. And,
finally, the presence of such quarks would have a very important
bearing on Higgs physics, both in terms of enhancing the glue-glue
fusion cross-section as well as through direct associated production
(with either of $t'$ and $b'$) rendering it quite important in the LHC
context. In other words, if the D0 anomaly stands the test of time and if 
a fourth generation is the explanation of the same, the experiments at the 
LHC would soon be in a position to validate it.
%%%%%%%%%%%%%%%%%%%%%%%%%%%%%%%%%%%%%%%%%%%%%%%%%%%%%%%%%%%%%%%%%%%%%%%%%%%%%
%\noindent
\section*{Acknowledgments}  
DKG acknowledges partial support from the Department of Science and
Technology, India under grant SR/S2/HEP-12/2006. The authors would like to
thank A. Kundu, A. Lenz for very useful discussions and 
the Theory Division, CERN for hospitality when this work was initiated.
%%%%%%%%%%%%%%%%%%%%%%%%%%%%%%%%%%%%%%%%%%%%%%%%%%%%%%%%%%%%%%
%\input{biblio.sect}
%%%%%%%%%%%%%%%%%%%%%%%%%%%%%%%%%%%%%%%%%%%%%%%%%%%%%%%%%%%%%%

%%%%%%%%%%%%%%%%%%%%%%%%%%%%%%%%%%%%%%%%%%%%%%%%%%%%%%%%%%%%%%
\end{document}